\documentclass[journal]{IEEEtran}
\usepackage{cite}
\ifCLASSINFOpdf
\usepackage[pdftex]{graphicx}
  \graphicspath{{../pdf/}{../jpeg/}}
  \DeclareGraphicsExtensions{.pdf,.jpeg,.png}
\else
  \usepackage[dvips]{graphicx}
  \graphicspath{{../eps/}}
  \DeclareGraphicsExtensions{.eps}
\fi
\usepackage{amsmath}
\usepackage{cases}
\usepackage{amsfonts}
\usepackage{subeqnarray}
\usepackage{subfigure}
\usepackage[ruled,linesnumbered]{algorithm2e}
\makeatletter
\newcommand{\nosemic}{\renewcommand{\@endalgocfline}{\relax}}
\newcommand{\dosemic}{\renewcommand{\@endalgocfline}{\algocf@endline}}
\let\oldnl\nl
\newcommand{\nonl}{\renewcommand{\nl}{\let\nl\oldnl}}
\makeatother
\usepackage[export]{adjustbox} 
\usepackage{booktabs}
\usepackage{setspace}
\usepackage{xcolor}
\usepackage{mathrsfs}
\usepackage{amsmath}
\usepackage{array}
\usepackage{amssymb}
\usepackage{amsthm}

\allowdisplaybreaks
\begin{document}
\setstretch{0.92}
\title{Privacy-Preserving Distributed Clustering for Electrical Load Profiling}
\author{Mengshuo~Jia, 
		Yi~Wang,
Chen~Shen 
      and Gabriela~Hug
\thanks{M. Jia and C. Shen are with the State Key Laboratory of Power Systems, Tsinghua University, 100084 Beijing, China. Y. Wang and G. Hug are with the Power Systems Laboratory, ETH Zurich, 8092 Zurich, Switzerland.}} 
\maketitle

\begin{abstract}
Electrical load profiling supports retailers and distribution network operators in having a better understanding of the consumption behavior of consumers. However, traditional clustering methods for load profiling are centralized and require access to all the smart meter data, thus causing privacy issues for consumers and retailers. To tackle this issue, we propose a privacy-preserving distributed clustering framework for load profiling by developing a privacy-preserving accelerated average consensus (PP-AAC) algorithm with proven convergence. Using the proposed framework, we modify several commonly used clustering methods, including k-means, fuzzy C-means, and Gaussian mixture model, to provide privacy-preserving distributed clustering methods. In this way, load profiling can be performed only by local calculations and information sharing between neighboring data owners without sacrificing privacy. Meanwhile, compared to traditional centralized clustering methods, the computational time consumed by each data owner is significantly reduced. The privacy and complexity of the proposed privacy-preserving distributed clustering framework are analyzed. The correctness, efficiency, effectiveness, and privacy-preserving feature of the proposed framework and the proposed PP-AAC algorithm are verified using a real-world Irish residential dataset.
\end{abstract}

\IEEEpeerreviewmaketitle


\section{Introduction}

\IEEEPARstart{A}{Massive} number of fine-grained electricity consumption data are being collected by smart meters. Identifying the load patterns from these smart meter data, i.e., residential load profiling, supports retailers and distribution network operators (DSO) in having a better understanding of the consumption behavior of consumers. For example, the retailers can provide personalized tariffs for different types of consumers; the DSO can perform detailed voltage simulation \cite{CLEMENTNYNS2011185} or micro-grid operation \cite{5720519} of the distribution network based on the identified load patterns.

Ideally, residential load profiling is carried out on a very large and diverse dataset to capture all different types of customers and behaviors. Particularly for retailers and third party providers such a diverse data set is important as they wish to design diversified electricity products to attract new consumers. However, residential load data are only monitored or collected by the corresponding retailers, i.e., each retailer only has the data of the consumers it serves. No center has access to all the smart meter data. Besides, since the smart meter data contains highly private information about the consumers \cite{5054916}, data sharing between retailers is not allowed. Thus, a privacy-preserving distributed clustering scheme is required, where retailers can possibly cooperate with others to jointly achieve the clustering results on their union consumption dataset via local calculation and communication. During the cooperation, the information of each retailer, e.g., the raw data or the number of consumers, will not be deduced by others.

So far, various clustering algorithms have been applied for load profiling, such as hierarchical clustering using different linkages \cite{4282059}, CFSFDP \cite{7448460}, k-means \cite{1626400}, fuzzy C-means algorithm (FCA) \cite{1432503}, Gaussian mixture model (GMM) \cite{6661463}, self organizing map \cite{RASANEN20103538}, etc. However, to the best of our knowledge, there is no relevant research on privacy-preserving distributed clustering for load profiling. 

To bridge this gap, this paper proposes a privacy-preserving distributed clustering framework for load profiling. This framework can be used to transform three commonly used clustering methods, i.e., k-means, FCA, and GMM, into distributed clustering algorithms for the purpose of privacy-preserving load profiling. Among these three methods, k-means is a `hard' clustering method that delivers deterministic clustering results \cite{LI20161530}; while FCA and GMM are `soft' methods that provide an extent or a probability measure of observations to each classification respectively, which can be leveraged to observe overlapping clusters or uncertain cluster memberships \cite{1056489}. 

In fact, many works about privacy-preserving clustering have been conducted in different fields such as marketing and medicine \cite{samet2007privacy}. Among them, the cryptography-based methods are most commonly used. These methods use secure multiparty computation \cite{manikandan2018privacy,clifton2002tools}, homomorphic encryption technique \cite{7903675,leemaqz2017corruption}, or the combination of both \cite{4221090} to turn the clustering methods into the privacy-preserving k-means \cite{7903675,4221090}, the privacy-preserving FCA \cite{manikandan2018privacy}, or the privacy-preserving GMM \cite{clifton2002tools,leemaqz2017corruption}. However, the methods using secure multiparty computation are extremely computationally expensive \cite{patel2012efficient}. Besides, the overheads of encryption in the homomorphic encryption technique also limit the scope of the corresponding clustering methods \cite{7847023} and result in time-consuming computations \cite{meskine2012privacy}. To reduce overheads, secret sharing can be adopted to design the privacy-preserving k-means clustering \cite{upmanyu2010efficient,patel2012efficient}. However, these secret-sharing-based methods, including the aforementioned cryptography-based methods, are not fully distributed algorithms, because each party (the data owners, like the retailers in this paper) either has to interact with a data center \cite{7903675,upmanyu2010efficient,manikandan2018privacy}, or has to communicate with all the other parties \cite{patel2012efficient,4221090}, or has to share its information along a pre-selected information transmission path \cite{clifton2002tools,leemaqz2017corruption,7847023}. These algorithms have the following drawbacks: (1) the existance of a data center or a preset information sharing path greatly increases the risk of a single point or single line failure; (2) the full communication between any two parties results in low scalability. 

The proposed privacy-preserving distributed clustering framework aims to solve the above issues. We first perform commonality analysis of the traditional k-means, FCA, and GMM, and point out that the key to the clustering framework lies in how to calculate the summation of retailers' private information in a fully distributed and privacy-preserving way. The average consensus (AC) algorithm, as an important fully distributed computing method in the automatic control area, provides the means to achieve the summation. However, the slow rate of its convergence towards the average is the major deficiency of this algorithm \cite{4694103}. Besides, the AC algorithm will reveal the private information available to the retailers during the interaction between neighbors. Therefore, we first introduce an accelerated AC (AAC) algorithm to significantly improve the rate of convergence without sacrificing the simplicity of the original AC algorithm \cite{4694103}. Then, we adapt the AAC algorithm to provide a privacy-preserving version by leveraging the exponentially decaying disturbance with zero-sum property proposed in \cite{8685131}. The convergence of the proposed privacy-preserving AAC (PP-AAC) algorithm is also proved. After that, we develop the privacy-preserving distributed clustering framework based on the proposed algorithm. This framework can convert the traditional k-means, FCA, and GMM into fully distributed privacy-preserving clustering methods, where each retailer only needs to communicate with its surrounding neighbors to obtain the exact load pattern identification results of all the consumers. Finally, we provide the privacy and complexity analyses of the proposed framework. 

This paper makes the following contributions:

\begin{itemize}
    \item Propose a privacy-preserving distributed clustering framework for load profiling. This framework is based on an original PP-AAC algorithm, which is theoretically proven to be convergent.
    \item Provide the privacy and complexity analyses of the proposed framework theoretically and practically. Results show that this framework not only protects the data privacy of retailers but also greatly reduces the computational overhead.
    \item Develop the privacy-preserving distributed k-means, FCA, and GMM clustering methods using the proposed framework. These methods are applied to identify electrical load patterns, whose results are the same as that of the centralized clustering methods. 
\end{itemize} 

To the best of our knowledge, this is the first time that the electrical load data has been analyzed using privacy-preserving distributed clustering methods.

The rest of this paper is organized as follows. Section II analyzes the commonality of k-means, FCA, and GMM. The PP-AAC algorithm is proposed in Section III.  Section IV develops the privacy-preserving distributed clustering framework for the three clustering methods. Case studies are provided in Section V, and Section VI concludes this paper.

\section{Problem Formulation}
This section first briefly reviews the standard clustering methods: k-means \cite{1056489}, FCA \cite{WU2012407}, and GMM \cite{5298967}, and then gives the commonality analyses of them. Before that, we assume that the union data set consists of $N$ observations. These observations are distributed among $M$ retailers, where retailer $i$ has $N_i$ consumers, i.e., $N_i$ observations. Besides, the centroid of cluster $k$, described by $\boldsymbol{\mu}_k$, is considered as the $k$-th load pattern of the union data set.

\subsection{K-means}

K-means partitions $N$ observations into $K$ clusters by minimizing the within-cluster variances as follows:
\begin{equation}
    \min\ \  f =\sum\nolimits_{k=1}^K \sum\nolimits_{i=1}^M \sum\nolimits_{n \in \boldsymbol{C}_k} \ \left\|\boldsymbol{y}_{i,n} - \boldsymbol{\mu}_k \right\|^2 \notag
\end{equation}
\noindent where $\boldsymbol{y}_{i,n}$ is the $n$-th observation of retailer $i$. $\boldsymbol{C}_k$ represents the index set of the observations belonging to cluster $k$. 
 
Although finding the solution is NP-hard, Lloyd’s algorithm guarantees to find a local minimum in a few iterations \cite{1056489}. First, $K$ initial cluster centroids are arbitrarily and randomly assigned. Then, in each iteration, the cluster index of $\boldsymbol{y}_{i,n}$ is computed by 
\begin{align}
	c_{i,n} :& = \mathop{\arg\min}_{\ \ \ \ \ k} \ \left\|\boldsymbol{y}_{i,n} - \boldsymbol{\mu}_k \right\| \label{k-means 1}
\end{align}
and the centroid of cluster $k$ is updated by 
\begin{align}
	\boldsymbol{\mu}_k & = \left. \sum\nolimits_{i=1}^M \boldsymbol{s}_{k,i} \middle/ \sum\nolimits_{i=1}^M \boldsymbol{z}_{k,i} \right. \label{k-means 2} \\
	\boldsymbol{s}_{k,i}  & = \sum\nolimits_{n=1}^{N_i} I(c_{i,n}=k)\ \boldsymbol{y}_{i,n} \label{k-means local 1} \\
	\boldsymbol{z}_{k,i} & = \sum\nolimits_{n=1}^{N_i} I(c_{i,n}=k) \label{k-means local 2}
\end{align}
sequentially. These two steps are repeated until convergence is achieved. Note that $I(a=b)$ equals $1$ if $a=b$ and $0$ otherwise.

\subsection{FCA}
FCA is the best-known method for fuzzy clustering with the objective function given as follows:
\begin{equation}
	\min \ \ f = \sum\nolimits_{k=1}^K \sum\nolimits_{i=1}^M \sum\nolimits_{n=1}^{N_i} \ \rho_{k,i,n}^m \left\|\boldsymbol{y}_{i,n} - \boldsymbol{\mu}_k \right\|^2 \notag
\end{equation}
where $m$ is the fuzziness index and $\rho_{k,i,n}^m$ is the degree to which $\boldsymbol{y}_{i,n}$ belongs to $\boldsymbol{C}_k$. The following iterative procedure solves this problem: the degree to which the observation belongs to cluster $k$ is first calculated by
\begin{align}
	\rho_{k,i,n}^m & = \frac{\left\|\boldsymbol{y}_{i,n} - \boldsymbol{\mu}_k \right\|^{-2/(m-1)}}{\sum\nolimits_{j=1}^K \left\|\boldsymbol{y}_{i,n} - \boldsymbol{\mu}_j \right\|^{-2/(m-1)}} \label{FCA 1}
\end{align}
Then, the centroid of cluster $k$ is updated by  
\begin{align}
	\boldsymbol{\mu}_k  & = \left. \sum\nolimits_{i=1}^M \boldsymbol{s}_{k,i} \middle/ \sum\nolimits_{i=1}^M \boldsymbol{z}_{k,i}  \right. \label{FCA 2} \\
	\boldsymbol{s}_{k,i}  & = \sum\nolimits_{n=1}^{N_i}\rho_{k,i,n}^m \boldsymbol{y}_{i,n} \label{FCA local 1} \\
	 \boldsymbol{z}_{k,i} & = \sum\nolimits_{n=1}^{N_i}\rho_{k,i,n}^m  \label{FCA local 2}
\end{align}
   
Different from k-means, where each observation either belongs to a cluster or not, FCA assigns degrees for each observation to be in every cluster, i.e., FCA is a type of soft clustering. 

\subsection{GMM}
As a convex combination of $K$ Gaussian components $\mathcal N_k$ with weight $\omega_k$ and covariance $\boldsymbol{\Sigma}_k$, GMM is given by
\begin{align}
	g(\boldsymbol{y}_{i,n}) = \sum\nolimits_{k=1}^K w_k \ 
	\mathcal N_k(\boldsymbol{y}_{i,n}|\boldsymbol{\mu}_k, \boldsymbol{\Sigma}_k) \label{GMM}  
\end{align}
where each Gaussian component represents a cluster. 

To divide the union data set into $K$ clusters by GMM, one should train GMM by leveraging the maximum likelihood estimation, which is given as follows:
\begin{align}
	\max \ \ & f = \prod_{k=1}^K\prod_{i=1}^M\prod_{n=1}^{N_i}\left[ \sum_{k=1}^K  w_k \ 
	\mathcal N_k(\boldsymbol{y}_{i,n}|\boldsymbol{\mu}_k, \boldsymbol{\Sigma}_k)\right] \notag \\
	\mbox{s.t.}\ \  & 0 \leq w_k \leq 1 , \ \ \sum\nolimits_{k=1}^K w_k = 1\notag 
\end{align}

The most commonly used maximum likelihood estimation method is the expectation-maximization (EM) algorithm \cite{5298967}, which can be summarized as two iterative steps: the E-step and the M-step. The E-step, as given in
\begin{equation}
    \label{E-step}
    Q_{k,i,n} = \frac{\omega_k\  \mathcal N(\boldsymbol{y}_{i,n}|\boldsymbol{\mu}_k,\boldsymbol{\Sigma}_k)}{\sum_{j=1}^K w_j\  \mathcal N(\boldsymbol{y}_{i,n}|\boldsymbol{\mu}_j,\boldsymbol{\Sigma}_j)} 
\end{equation}
computes the probability that an observation belongs to cluster $k$. The M-step updates the parameters in (\ref{GMM}) according to 
\begin{align}
	\omega_k & = \frac{1}{N} \sum\nolimits_{i=1}^{M}\boldsymbol{z}_{k,i} 	\label{M-step 1}  \\
	\boldsymbol{\mu}_k & = \left.\sum\nolimits_{i=1}^{M}\boldsymbol{s}_{k,i} \middle/ \sum\nolimits_{i=1}^{M}\boldsymbol{z}_{k,i}\right. \label{M-step 2} \\
	\boldsymbol{\Sigma}_k & = \left.\sum\nolimits_{i=1}^{M}\boldsymbol{h}_{k,i} \middle/ \sum\nolimits_{i=1}^{M}\boldsymbol{z}_{k,i} \right.  \label{M-step 3} \\
	\boldsymbol{s}_{k,i} & = \sum\nolimits_{n=1}^{N_i} Q_{k,i,n} \boldsymbol{y}_{i,n} \label{M-step local 1} \\
	\boldsymbol{z}_{k,i} &  = \sum\nolimits_{n=1}^{N_i} Q_{k,i,n}  \label{M-step local 2} \\
	 \boldsymbol{h}_{k,i} & = \sum\nolimits_{n=1}^{N_i} Q_{k,i,n} (\boldsymbol{y}_{i,n}-\boldsymbol{\mu}_k)^T(\boldsymbol{y}_{i,n}-\boldsymbol{\mu}_k) \label{M-step local 3}
\end{align}

After convergence, the final parameters of (\ref{GMM}) are reached. The final probability that an observation belongs to cluster $k$ can be obtained by substituting the final parameters into (\ref{E-step}). Same as FCA, GMM is also a soft clustering method.

\subsection{Commonality Analysis}
The clustering of k-means, FCA, and GMM have two points in common, which are listed in Remark 2.1 and 2.2.

\textit{\textbf{Remark 2.1}}: The clustering processes of k-means, FCA, and GMM can all be summarized in two parts: the local calculation part and the global calculation part, where the local one can be performed by each retailer, and the global one is essentially the summation of each retailer's local calculation results.

In fact, each retailer can directly perform the first steps of the three algorithms via its own data, i.e., the calculation in (\ref{k-means 1}), (\ref{FCA 1}) or (\ref{E-step}). Then, retailer $i$ is able to compute the following local results $\boldsymbol{L}_{k,i}$:
\begin{equation}
	\label{local}
	\boldsymbol{L}_{k,i} =\left\{
	\begin{split}
	& \boldsymbol{s}_{k,i}\  \text{in}\  (\ref{k-means local 1}),\  \boldsymbol{z}_{k,i} \ \text{in}\  (\ref{k-means local 2}),\ \text{for k-means} \\
	& \boldsymbol{s}_{k,i}\  \text{in}\  (\ref{FCA local 1}),\  \boldsymbol{z}_{k,i} \ \text{in}\  (\ref{FCA local 2}),\ \text{for FCA} \\
	& \boldsymbol{s}_{k,i}\  \text{in}\  (\ref{M-step local 1}),\  \boldsymbol{z}_{k,i} \ \text{in}\  (\ref{M-step local 2}),\  \boldsymbol{h}_{k,i} \ \text{in}\  (\ref{M-step local 3}),\ \text{for GMM} \\
	\end{split} \right.
\end{equation}
depending on the algorithm used. Once each retailer obtains the local results, the global summation of those local results from all retailers is required to continue the clustering method. For example, k-means algorithm needs to sum the local results $\boldsymbol{s}_{k,i}$ and $\boldsymbol{z}_{k,i}$ of all retailers respectively to update the centroid of cluster $k$ in (\ref{k-means 2}). Let $\boldsymbol{G}_k$ be the global summation result, then we have:
\begin{equation}
	\label{global}
	\boldsymbol{G}_k =\left\{
	\begin{split}
	& \sum\nolimits_{i=1}^M \boldsymbol{s}_{k,i},\  \sum\nolimits_{i=1}^M\boldsymbol{z}_{k,i} \ \text{in}\  (\ref{k-means 2}),\ \text{for k-means} \\
	& \sum\nolimits_{i=1}^M \boldsymbol{s}_{k,i},\  \sum\nolimits_{i=1}^M\boldsymbol{z}_{k,i} \ \text{in}\  (\ref{FCA 2}),\ \text{for FCA} \\
	& \sum\nolimits_{i=1}^M \boldsymbol{s}_{k,i},\  \sum\nolimits_{i=1}^M\boldsymbol{z}_{k,i},\  \sum\nolimits_{i=1}^M\boldsymbol{h}_{k,i}\ \text{in}\  (\ref{M-step 1})\text{-}(\ref{M-step 3})\\
	& \qquad \qquad \qquad \qquad \qquad \qquad \qquad \qquad \ \ \text{for GMM}
	\end{split} \right.
\end{equation}
Therefore, the relationship between the local and the global calculation parts can be generalized to:
\begin{equation}
	\boldsymbol{G}_k = \sum\nolimits_{i=1}^M \boldsymbol{L}_{k,i}
\end{equation}
where $\boldsymbol{L}_{k,i}$ can be calculated by each retailer locally using (\ref{local}), while the computation of $\boldsymbol{G}_k$ needs cooperations among all retailers. Once $\boldsymbol{G}_k$ in (\ref{global}) is obtained, the second steps of the three algorithms can be carried out and the iterative procedure continues. 

\textit{\textbf{Remark 2.2}}: Each retailer's local calculation results from k-means, FCA, and GMM contain private information, so that retailer $i$ will refuse to share its $\boldsymbol{L}_{k,i}$ with others.

In fact, if retailer $i$ shares its $\boldsymbol{L}_{k,i}$ ($\forall k$) with retailer $j$, the latter can derive the following private information of retailer $i$:

\subsubsection{The number of retailer $i$'s consumers}

Once retailer $j$ has received $\boldsymbol{z}_{k,i}$ ($\forall k$) in (\ref{k-means local 2}), (\ref{FCA local 2}) or (\ref{M-step local 2}), it can compute the number via:
\begin{equation}
	N_i = \sum\nolimits_{k=1}^K \boldsymbol{z}_{k,i} \notag
\end{equation}

\subsubsection{The proportion or number of retailer $i$'s consumers belonging to cluster $k$}

Once retailer $j$ receives $N_i$, it will also obtain the proportion of retailer $i$'s consumers belonging to cluster $k$ by:
\begin{equation}
	r_{k,i} = \left. \boldsymbol{z}_{k,i} \middle/ N_i \right. \notag
\end{equation}
Particularly, retailer $j$ can directly know the specific number of retailer $i$'s consumers belonging to cluster $k$ by receiving $\boldsymbol{z}_{k,i}$ in (\ref{k-means local 2}).

\subsubsection{Retailer $i$'s local load pattern of cluster $k$}

Once retailer $j$ has received $\boldsymbol{s}_{k,i}$ in (\ref{k-means local 1}), (\ref{FCA local 1}) or (\ref{M-step local 1}), along with $\boldsymbol{z}_{k,i}$ in hand, retailer $j$ can compute the local centroid of retailer $i$ in cluster $k$ by: 
\begin{equation}
	\boldsymbol{\mu}_{k,i} = \left. \boldsymbol{s}_{k,i} \middle/ \boldsymbol{z}_{k,i} \right. \notag
\end{equation}
which will reveal the approximate load pattern of retailer $i$. For example, we choose $\boldsymbol{s}_{k,i}$ in (\ref{k-means local 1}) and $\boldsymbol{z}_{k,i}$ in (\ref{k-means local 2}), then $\boldsymbol{\mu}_{k,i}$ is essentially the mean of retailer $i$'s observations belonging to cluster $k$, which can be considered as its approximate load pattern in cluster $k$. The approximation lies in the fact that $\boldsymbol{s}_{k,i}$ and $\boldsymbol{z}_{k,i}$ are calculated using the global centroid in (\ref{k-means 2}) in the last iteration, not the local centroid of retailer $i$ in the last iteration; otherwise it will be the exact load pattern based on retailer $i$'s data set.

\textit{\textbf{Definition 2.3}}: We define the ``privacy'' of retailer $i$ ($\forall i$) as the information set $\boldsymbol{P}_i =\{N_i, r_{k,i}, \boldsymbol{\mu}_{k,i}|k=1,...,K\} $.

Clearly, retailer $i$ will not let retailer $j$ obtain $\boldsymbol{P}_i$. As a result, directly sharing $\boldsymbol{L}_{k,i}$ will be refused by retailer $i$, impeding the implementation of the key summation in (\ref{global}) for the three algorithms. Therefore, a privacy-preserving distributed summation algorithm to compute (\ref{global}) is required.

\section{PP-AAC Algorithm}
To achieve a distributed summation algorithm, this section first introduces an AAC algorithm with a fast convergence rate \cite{4694103}. After that, we further improve the AAC algorithm by leveraging an exponentially decaying disturbance with zero-sum property to propose a PP-AAC algorithm. Finally, the convergence of the proposed algorithm is proved.

\subsection{AAC Algorithm}
The AAC algorithm is graph-theory-based. Therefore, we consider a graph consisting of the $M$ nodes and $n_l$ edges. Each node represents a retailer, and the edge between each pair of nodes means that there is bidirectional noise-free communication between two retailers. This graph is publicly known by all retailers. Denote the node set by $\boldsymbol{\mathcal V}$ and the edge set by $\boldsymbol{\varepsilon}$. The neighborhood of retailer $i$ is represented by $\boldsymbol{\Omega}_i \triangleq \{j\in\boldsymbol{\mathcal V}: \{i,j\} \in \boldsymbol{\varepsilon}\}$, and the degree of retailer $i$ is denoted by $d_i$. Let $\boldsymbol{W} \in \Re^{M \times M} $ be the Metropolis weight matrix with elements as follows \cite{1440896}:
\begin{equation}
	\label{weight matrix}
	W_{ij} =\left\{
	\begin{split}
	& \frac{1}{1+\max\{d_i,d_j\} }\quad \text{if $j \in \boldsymbol{\Omega}_i $} \\
	& 1-\sum\nolimits_{j \in \boldsymbol{\Omega}_i} W_{i,j} \quad \text{if $i=j$} \\
	& 0 \qquad\qquad\quad\quad\quad\ \  \text{Otherwise} \\
	\end{split} \right.
\end{equation}

In the AAC algorithm, each retailer has a state value that will be updated through iterations. Let $x_i$ be the state of retailer $i$ in the AAC algorithm, then the state update equation of the AAC algorithm in the $t$-th iteration is given by
\begin{align}
	x_i(t+1) & = \alpha x_i^p(t+1) + (1-\alpha) x_i^w(t+1) \label{AAC algorithm}
\end{align}
which is a convex combination of the value from the original AC algorithm and the predictor given respectively by
\begin{align}
	x_i^w(t+1) & = W_{i,i}x_i(t) + \sum\nolimits_{j \in \boldsymbol{\Omega}_i } W_{i,j}x_j(t) \label{ACA} \\
	x_i^p(t+1) & = 2\cdot x_i^w(t+1) - x_i(t) \label{predictor}
\end{align}
The matrix form of the update is given as follows: 
\begin{align}
	& \boldsymbol{W}^\ast \triangleq  (1+\alpha)  \boldsymbol{W} - \alpha \boldsymbol{I} \label{W star} \\
	& \boldsymbol{X}(t+1) = \boldsymbol{W}^\ast \boldsymbol{X}(t) \label{AAC algorithm matrix form} 
\end{align}
where $\boldsymbol{X}(t) = [x_1(t),...,x_M(t)]^T$, and $\boldsymbol{I} \in \Re^{M \times M}$ is the identity matrix. We call $\boldsymbol{W}^\ast $ the accelerated Metropolis weight matrix.

In this way, $x_i$ will converge to the mean of all retailers' initial standardized state values
\begin{equation}
	\lim\limits_{t \to \infty }{x_i(t)} = \frac{1}{M} \sum\nolimits_{i=1}^M x_i(0) \label{mean} 
\end{equation}
with the fastest asymptotic worst-case convergence rate if the weighted coefficient $\alpha$ equals the optimal value \cite{4694103}:
\begin{equation}
	\alpha = \frac{\lambda_M+\lambda_2}{2-\lambda_M-\lambda_2} 
\end{equation}
where $\lambda_M$ is the smallest eigenvalue of $\boldsymbol{W}$, and $\lambda_2$ is the second largest eigenvalue of $\boldsymbol{W}$. Since the graph is publicly known by all retailers, each retailer can easily compute $\boldsymbol{W}$ using \eqref{weight matrix}. Then, $\boldsymbol{W}^\ast$ can be obtained by all retailers using \eqref{W star}.

Note that the AAC algorithm is fully distributed, i.e., each retailer only needs to communicate with its neighbors. Besides, after convergence, retailers can obtain the summation of their initial standardized state values by multiplying the mean in (\ref{mean}) by $M$. Thus, let $x_i(0)$ be equal to $\boldsymbol{L}_{k,i}$, then each retailer can obtain $\boldsymbol{G_k}$ in (\ref{global}) in a fully distributed manner using the AAC algorithm. However, in the first iteration, retailer $i$ will send $x_i(0) = \boldsymbol{L}_{k,i} $ to its neighbors, which directly reveals the private information of retailer $i$. 

\subsection{PP-AAC Algorithm}

To facilitate the AAC algorithm with privacy-persevering characteristics, we utilize the exponentially decaying disturbance with zero-sum property from \cite{8685131} to mask the interactive state values among neighbors during the AAC iterations, so that each retailer cannot derive private information of the others.

The proposed PP-AAC algorithm is defined by  
\begin{equation}
	x_i(t+1) = W_{i,i}^\ast x_i^+(t) + \sum\nolimits_{j \in \boldsymbol{\Omega}_i } W_{i,j}^\ast x_j^+ (t)  \label{PP AAC algorithm}
\end{equation}
where $x_i^+(t)$ is the state value masked by the disturbance $\theta_i(t)$ as follows:
\begin{equation}
	\begin{split}
		& x_i^+(t) = x_i(t) + \theta_i(t) \\
		& \theta_i(t) = \delta_i(t) - \delta_i(t-1) \label{mask}
	\end{split}
\end{equation}

\noindent The noise
$\delta_i(t)$ is randomly selected from $[-\frac{\sigma}{2}\beta^{t+1}, \frac{\sigma}{2}\beta^{t+1}]$ by retailer $i$, where $\sigma>0$, $\beta \in [0,1)$, and $\delta(t<0) = 0$. This design leads to the two features of $\theta_i(t)$, which will be used for the following proof of Theorem 3.1:  
\begin{itemize}
	\item The noise $\delta_i(t)$ is exponentially decaying as $\beta \in [0,1)$ and $t$ grows with the number of iterations. So $\theta_i(t)$ is also exponentially decaying.
	\item The disturbance $\theta_i(t)$ has zero-sum property, which means that if we sum up $\theta_i(t) \ (\forall i)$ from $t=0$ to infinity (or to a relatively large number), the result will be close to 0, i.e.,
\end{itemize}
\begin{equation}
	\sum\nolimits_{i=1}^M \sum\nolimits_{t=0}^\infty \theta_i(t) = \sum\nolimits_{i=1}^M \lim\limits_{t \to \infty }  {\delta_i(t)} \rightarrow 0 \label{zero-sum}
\end{equation}

\textit{\textbf{Theorem 3.1}}: The proposed PP-AAC algorithm in (\ref{PP AAC algorithm}) will make each retailer's state value converge to the average of all retailers' initial state values, i.e., (\ref{mean}) still holds.

\textit{\textbf{Proof}}: See Appendix.

\section{Privacy-preserving Distributed Clustering Framework}

This section describes the privacy-preserving distributed clustering framework for k-means, FCA, and GMM incorporating the proposed PP-AAC algorithm. In addition, we provide the privacy and complexity analyses of the proposed framework.

\subsection{Clustering Framework}

The idea of the clustering framework is that independent of the employed clustering method, in every iteration, each retailer first performs its local calculation according to (\ref{local}); then each retailer sets its local result as the initial state of the proposed PP-AAC algorithm; after convergence, each retailer obtains the global summation of all the local results in (\ref{global}); finally, using the global summations, each retailer can perform the rest of the clustering method to update the global information, e.g., the centroids of all clusters. The detailed clustering framework is demonstrated in Algorithm \ref{clustering framework}.

\begin{algorithm}
	\label{clustering framework}
	\caption{The clustering framework}
	\KwIn{Standardized $\boldsymbol{y}_{i,n}$ ($n=1,...,N_i$) of retailer $i$ ($\forall i$).}
	\KwIn{Arbitrarily and publicly assign K centroids $\boldsymbol{\mu}_k$.}
    \KwOut{The load pattern $\boldsymbol{\mu}_k$ ($\forall k$) of the union data set}
    \While{convergence criterion of clustering is not met}
	{
		Retailer $i$ ($\forall i$) calculates $\boldsymbol{L}_{k,i}$ ($\forall k$) in (\ref{local})\;
		Retailer $i$ ($\forall i$) sets $\boldsymbol{x}_i(0) = \left[\boldsymbol{L}_{1,i},...,\boldsymbol{L}_{K,i} \right] $\;
		$t=0$\;
		\While{average consensus is not achieved}
		{
			Retailer $i$ ($\forall i$) randomly selects $\delta_i(t)$ by rule\;
			Retailer $i$ ($\forall i$) masks its $\boldsymbol{x}_i(t)$ by (\ref{mask})\;
			Retailer $i$ ($\forall i$) computes its $\boldsymbol{x}_i(t+1)$ by (\ref{PP AAC algorithm})\;
			$t=t+1$\;
		}
		Retailer $i$ ($\forall i$) obtains $\left[\boldsymbol{G}_{1,i},...,\boldsymbol{G}_{K,i} \right] $ by $M\times \boldsymbol{x}_i(t)$\;
		Retailer $i$ ($\forall i$) updates global cluster information\;
		$\quad$ - K-means: updates $\boldsymbol{\mu}_k$ ($\forall k$) by (\ref{k-means 2})\;	
		$\quad$ - FCA: updates $\boldsymbol{\mu}_k$ ($\forall k$) by (\ref{FCA 2})\;
		$\quad$ - GMM: updates $\omega_k, \boldsymbol{\mu}_k, \boldsymbol{\Sigma}_k$ ($\forall k$) by (\ref{M-step 1})-(\ref{M-step 3})\;	
	}   
	Retailer $i$ ($\forall i$) gets the load patterns of the union data set\;
	$\quad$ - K-means: gets the final $\boldsymbol{\mu}_k$ ($\forall k$) in (\ref{k-means 2})\;	
	$\quad$ - FCA: gets the final $\boldsymbol{\mu}_k$ ($\forall k$) in (\ref{FCA 2})\;
	$\quad$ - GMM: gets the final $\boldsymbol{\mu}_k$ ($\forall k$) in (\ref{M-step 2})\;	
\end{algorithm}

\vspace{-0.3cm}
\subsection{Privacy Analysis}
As aforementioned, the AAC algorithm will directly reveal the initial value $\boldsymbol{x}_j(0)$ in the first iteration. On the contrary, in the first iteration of the proposed PP-AAC algorithm, retailer $i$ ($\forall i$) receives $\boldsymbol{x}_j^+(0)$ ($\forall j \in \boldsymbol{\Omega}_i$) instead of $\boldsymbol{x}_j(0)$. Since $\boldsymbol{x}_j^+(0)$ is masked using independent disturbance $\theta_j(0)$ by retailer $j$, retailer $i$ cannot derive the original value of $\boldsymbol{x}_j(0)$ from $\boldsymbol{x}_j^+(t)$, thus retailer $i$ will not know the private $\boldsymbol{L}_{k,j}$ of its neighbors, protecting the private information $\boldsymbol{P}_j$ of retailer $j$. In the remaining iterations, the process of adding disturbance continues; meanwhile, $\boldsymbol{x}_j(t)$ begins to converge to the mean value in (\ref{mean}) and moves away from its initial value, which further masks the true initial value. Quantitative illustrations will be shown in the next section.

In addition, we should note that if $j \in \boldsymbol{\Omega}_i$ and $\boldsymbol{\Omega}_j \subseteq  \boldsymbol{\Omega}_i$, i.e., retailer $i$ can receive all the information that retailer $j$ has received, including retailer $j$'s information, then retailer $i$ can deduce retailer $j$'s initial value even if the disturbance is introduced \cite{8685131}. Therefore, the authors in \cite{8685131} and \cite{7465717} both consider it necessary to assume that retailer $i$ cannot receive all the information that retailer $j$ has. The assumption is also adopted in this paper. Since $\boldsymbol{W}$ is publicly known by all retailers, retailer $j$ can tell that whether $\boldsymbol{\Omega}_j$ is a subset of its neighbor's $\boldsymbol{\Omega}_i$. If such a situation occurs, retailer $j$ can refuse to communicate with retailer $i$. Therefore, the assumption will hold in practice. 

\subsection{Complexity Analysis}
For the distributed framework, we investigate each retailer's computation and communication overhead. 

The proposed clustering framework not only keeps all the multiplication calculations in the original clustering methods, but also introduces new multiplication calculations by integrating the proposed PP-AAC algorithm. The multiplication calculations in the original clustering methods are divided by retailers according to their number of observations, i.e., if the computation overhead of the original clustering method is $\mathcal O(\varphi) $, then the overhead of retailer $i$ is $\mathcal O(\varphi N_i/N) $. Moreover, in each iteration of the PP-AAC algorithm, although the disturbance can be queried from the preset lookup table, retailer $i$ ($\forall i$) still needs to compute $W^\ast_{i,i}x_i^+(t)$ and $W^\ast_{i,j}x_j^+(t)$ ($\forall j \in \boldsymbol{\Omega}_i$), which requires $\widetilde{d}_i = d_i+1$ multiplications. Let $T_c$ denote the iteration number of the selected clustering method, and $T_a$ represent the iteration number of the proposed AAC algorithm, then the computation overhead of retailer $i$ is $\mathcal O(\varphi N_i/N + \widetilde{d}_iT_aT_c) $. Take k-means for example, where $\varphi$ is $NKT_c$, then retailer $i$'s overhead is $\mathcal O(N_iKT_c + \widetilde{d}_iT_aT_c) $. Please note that $\widetilde{d}_iT_a\ll NK$, because the number of retailers in a DN is small, and the proposed PP-AAC algorithm's convergence is accelerated, thus $T_a$ is generally also small. However, $N$ is thousands and $K \geq 2$. Moreover, we know that $N_i\ll N$. Therefore, the computation overhead of retailer $i$ is significantly smaller than that of the centralized k-means. Detailed illustrations is shown in the next section.

Besides, in each iteration of the proposed AAC algorithm, the communication number of retailer $i$ is $d_i$ \cite{MO2010209}. Therefore, the communication overhead of retailer $i$ is $\mathcal O(d_i T_aT_c) $.

\section{Case Study}

\subsection{Data Description and Experiment Setup}
\begin{figure}[h]
	\centering  
	\includegraphics[width=1.5in,center]{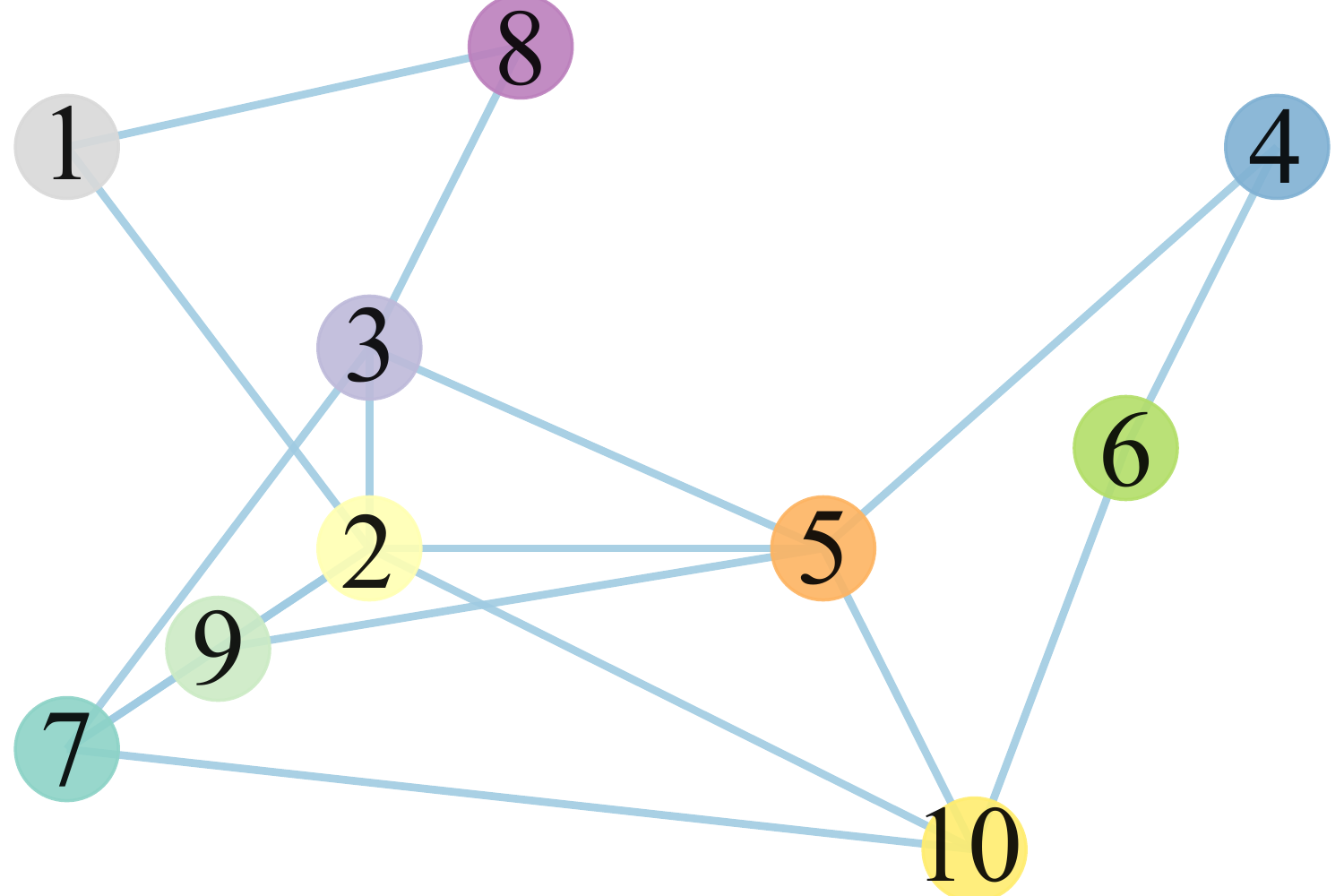}
	\caption{Communication topology of the retailers.}
	\label{case_topo}
\end{figure}

We utilize the smart meter data from Ireland for verification, which contains 509660 half-hourly daily electrical consumption observations of 1000 consumers \cite{Irish}. The representative load profile (RLP) of each consumer is obtained via the method presented in \cite{7579208}. Thus we get the union data set consisting of 1000 48-dimensional RLPs. For the verification of the proposed PP-AAC algorithm and the clustering framework, e.g., the correctness, the efficiency, the privacy-preserving feature, and the effectiveness, we assume that there are 10 retailers in a DN, and each of them has access to 100 consumers. Their initial communication topology is shown in Fig. \ref{case_topo}, where each retailer only communicates with its one-hop neighbors, and retailer $i$ ($\forall i$) cannot receive all the information that any of its neighbors has. We also use different topologies to investigate the trend of the computational cost of the proposed clustering framework with respect to different topologies. Besides, we set $\sigma=2$ and $\beta=0.2$ for randomly selecting the disturbance. Meanwhile, the initial centroids for all clustering methods are randomly chosen. 

\subsection{Verification of the PP-AAC algorithm}

\begin{figure}[h]
	\centering
	\includegraphics[width=2.0in,center]{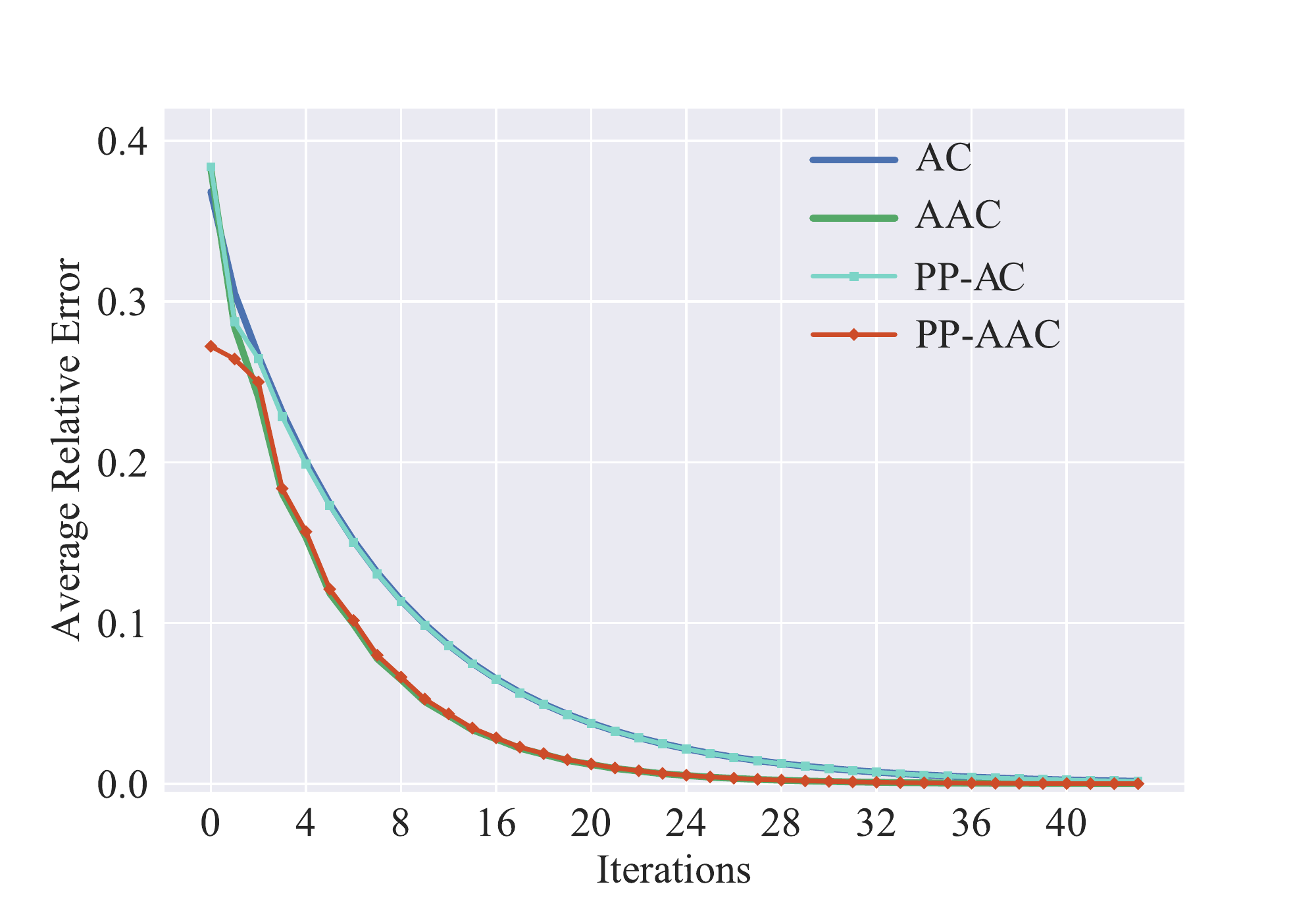}
	\caption{Comparison of convergence of the average consensus approaches and their respective privacy-preserving distributed versions.}
	\label{case_iter}
\end{figure}

To verify the correctness and efficiency of the proposed PP-AAC algorithm, we compare it with three algorithms: the original AC algorithm in \cite{1440896}, the AAC algorithm proposed in \cite{4694103} and the PP-AC algorithm proposed in \cite{8685131}. We use the four algorithms to compute the summation of the observations from each retailer's first consumer. We then illustrate the average error of all retailers relative to the accurate summation result. The errors of the four algorithms for each iteration are shown in Fig. \ref{case_iter}. It can be observed that the average error of the proposed PP-AAC algorithm converges to 0, indicating the correctness of this method. In addition, the proposed algorithm has the same convergence rate as the AAC algorithm. The PP-AC algorithm also has the same convergence rate as the AC algorithm. Please note that the proposed algorithm converges faster than both the AC and the PP-AC algorithm, indicating the efficiency of the proposed algorithm. Therefore, the correctness and efficiency of the proposed PP-AAC algorithm are verified.
\vspace{-0.3cm} 

\begin{figure}[h]
	\centering
	\includegraphics[width=2.1in,center]{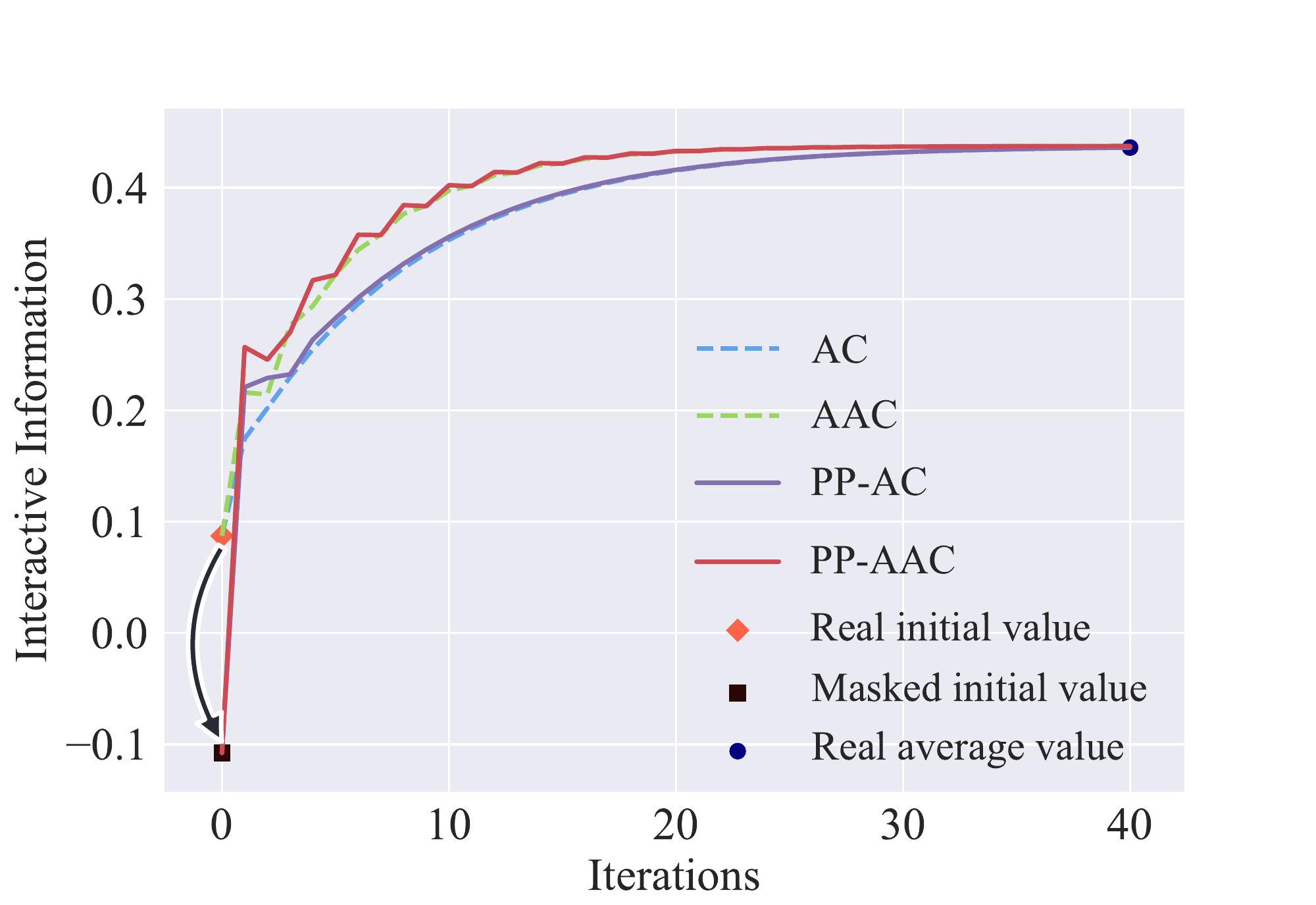}
	\caption{Interactive information shared during the iterative process.}
	\label{case_mask_value}
\end{figure}

Compared to the AC algorithm and the AAC algorithm, the proposed algorithm also has the privacy-preserving feature. To illustrate this feature, we provide the value that retailer 1 shares with its neighbors during the above summation calculation at each iteration. The shared values of the four algorithms are shown in Fig. \ref{case_mask_value}. 
\vspace{-0.3cm}

These shared values all converge to the real average value, but we should note that retailer 1 shares its real initial value with its neighbors in the first iteration when performing the AC and the AAC algorithm, which directly reveals the private information of retailer 1. However, after introducing the disturbance for masking, the proposed algorithm enables retailer 1 to share its masked initial value to its neighbors, which is far away from the real one as indicated by the black arrow. Thus, the proposed algorithm protects the privacy of retailer 1. Moreover, the proposed algorithm still converges faster than the PP-AC algorithm, even if they both start from the same masked initial point. 

\subsection{Verification of the Proposed clustering framework}

We can employ the proposed clustering framework to obtain privacy-preserving distributed k-means, FCA, and GMM clustering methods. Then, we use them for load pattern identification on the distributed data sets. As benchmarks, we also use the centralized k-means, FCA, and GMM for load pattern identification on the corresponding union data set.

\begin{figure}[h]
	\centering  
	\includegraphics[width=3in,center]{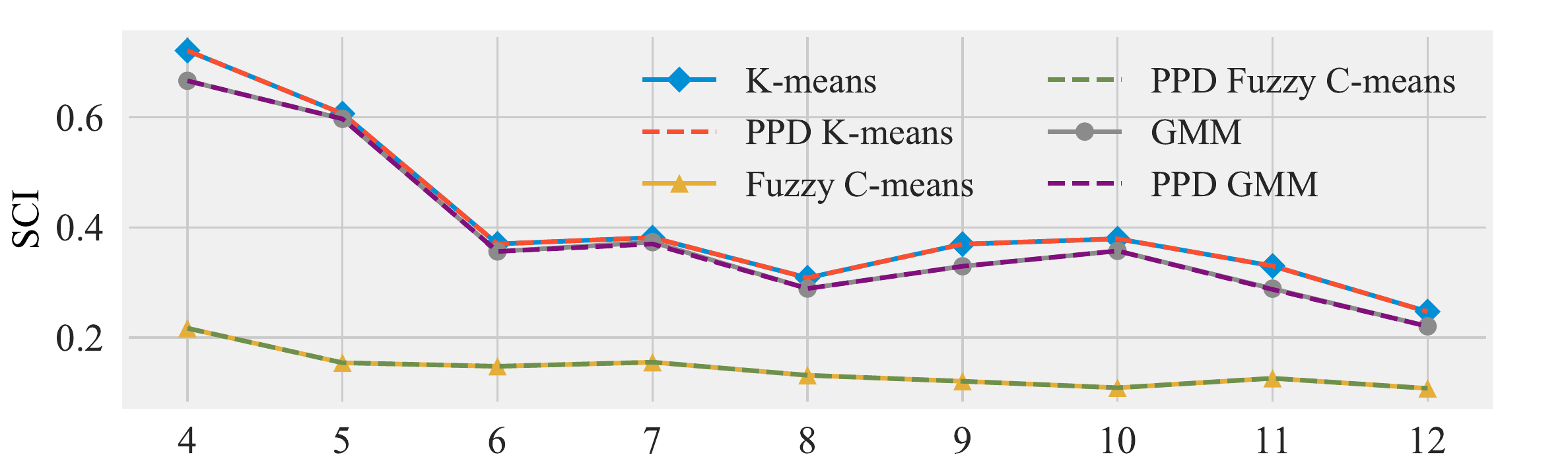}
	\caption{The indicators results of different clustering methods and their respective privacy-preserving distributed versions.}
	\label{case_indic}
\end{figure}

To verify the correctness of the clustering framework, in Fig. \ref{case_indic}, we use the Silhouette coefficient index (SCI) \cite{LLETI200487} to evaluate the above distributed and centralized algorithms for a different numbers of clusters. Note that the abbreviation `PPD' in Fig. \ref{case_indic} represents `privacy-preserving distributed'. This figure clearly shows that the SCI results of the proposed privacy-preserving distributed algorithms are identical to those of the centralized algorithms. This means that the clustering results on the distributed data sets using the proposed clustering framework, are exactly the same as those on the union data set computed via the centralized methods, indicating the correctness of the proposed clustering framework. 
\vspace{-0.4cm}

\begin{figure}[h]
	\centering  
	\includegraphics[width=3.5in,center]{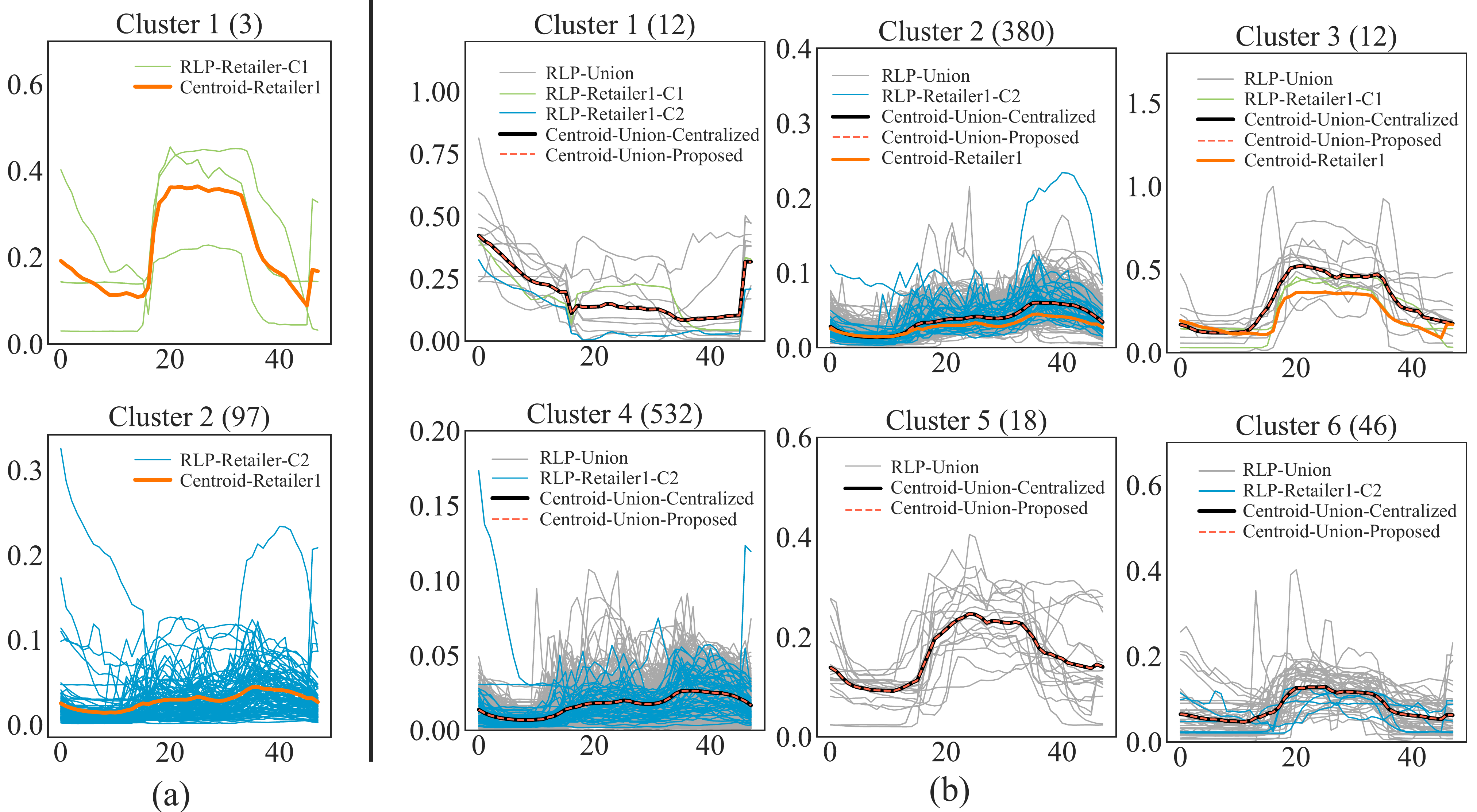}
	\caption{The clustering results on (a) the  data set of retailer 1 (100 RLPs) and (b) the union data set (1000 RLPs). The RLPs and clustering centers of retailer 1 are also illustrated in (b).}
	\label{case_cluster}
\end{figure}

To verify the effectiveness of the clustering framework, we choose k-means for demonstration as it is a hard clustering method, which is very convenient for illustration. We use the most common way, i.e., the sum of squared errors (SSE), to find the optimal number of clusters \cite{KWEDLO20111613}. From this, we find that the optimal cluster number of the union data set (1000 RLPs, i.e., 1000 consumers) is $K=6$, while that of the data set of retailer 1 (100 RLPs) is $K=2$. After that, we perform the centralized k-means on retailer 1's data set, and the results are shown in Fig. \ref{case_cluster}(a). Besides, we also perform the proposed privacy-preserving distributed k-means and the centralized k-means on the union data set. The results are demonstrated in Fig. \ref{case_cluster}(b). The number of RLPs in each cluster is listed in the sub figure's title. Meanwhile, the RLPs and the load patterns of retailer 1 are highlighted in Fig. \ref{case_cluster}(b) as well. 

First, from Fig. \ref{case_cluster}(b), we can observe that the centroids of the proposed algorithm are coincident with the centroids of the centralized k-means. Second, the two load patterns of retailer 1's data set in Fig. \ref{case_cluster}(a), approximately match the 2nd and the 3rd load patterns of the union data set in Fig. \ref{case_cluster}(b). However, retailer 1 missed the remaining four categories of consumers. Certainly, if retailer 1 only uses its own two load patterns for tariff design, its products will be difficult to attract the 608 consumers in the remaining clusters. On the contrary, by the proposed clustering framework, each retailer can use the six load patterns of all consumers for tariff design to attract all of them. Therefore, the effectiveness of the proposed clustering framework is proven.

To verify the efficiency of the clustering framework, we provide the computational time and iteration numbers (I-Ns) of the centralized and the privacy-preserving distributed clustering methods. Note that for the distributed methods, the retailers' computational times are different. Thus the maximum computational time of all retailers is chosen to represent the time of the distributed methods. Details are given in Table \ref{Cost_fixed_topo}. From this table, it is obvious that the iteration numbers of the corresponding centralized and distributed clustering methods are the same, but the computational times of the corresponding methods differ by an order of magnitude: the time consumed by each retailer in distributed clustering is significantly less than that of the centralized clustering, indicating the high efficiency of the proposed clustering framework.
\vspace{-0.45cm}
\begin{table}[h]
\setlength{\abovecaptionskip}{0pt}
	\renewcommand{\arraystretch}{1.1}
	\caption{Computational Time and Iteration Number Comparison}
	\label{Cost_fixed_topo}
	\centering
	\footnotesize
	\setlength{\tabcolsep}{0.9mm}{
	\begin{tabular}{c c c c c c c}
	\hline
	\bfseries Methods & K-means & PPD K-means & FCA & PPD FCA & GMM & PPD GMM \\
	\hline
	 \bfseries Time & 0.321   & 0.046 & 1.169 & 0.163  & 18.698 & 2.081 \\
	 \bfseries I-N & 7 & 7 & 24 & 24 & 6 & 6\\
	\hline
	\end{tabular}}
\end{table}

Please note that the above computational time does not contain communication time. However, this time is probably negligible. In k-means for example, each retailer shares its masked value to its neighbors, which consists of the masked $\boldsymbol{s}_{k,i} \in \Re^{48 \times 1} $ and $\boldsymbol{z}_{k,i} \in \Re^{1 \times 1} $ for $k=1,...,6$. Thus each retailer actually shares 294 floating-point numbers with its neighbors, i.e., 1.15 kbytes. We know that $N=1000$, $N_i=100$, $K=6$ and $T_c=7$. Meanwhile, $T_a=27$ as shown in Fig. \ref{case_iter}, and the degree of the retailer that consumes the most time is $5$ (retailer 1), which is also the maximum degree among the retailers. According to the communication overhead analysis in Section IV-C, the maximum total amount of upstream data of all the retailers will be $1.15 \times 5 \times 27 \times 7 = 1086.75$ kbytes $\approx 1.06$ Mbytes. Since the global average broadband internet speed is 11.03Mbps, the actual maximum communication time for retailers will not exceed 0.1 seconds. This cost will be greatly reduced in Europe as it has the world’s highest concentration of countries with the fastest internet, e.g., Sweden's average speed is 55.18Mbps \cite{internet}.

\subsection{Verification of Different Topologies}

\begin{figure}[h]
	\centering  
	\includegraphics[width=2.7in,center]{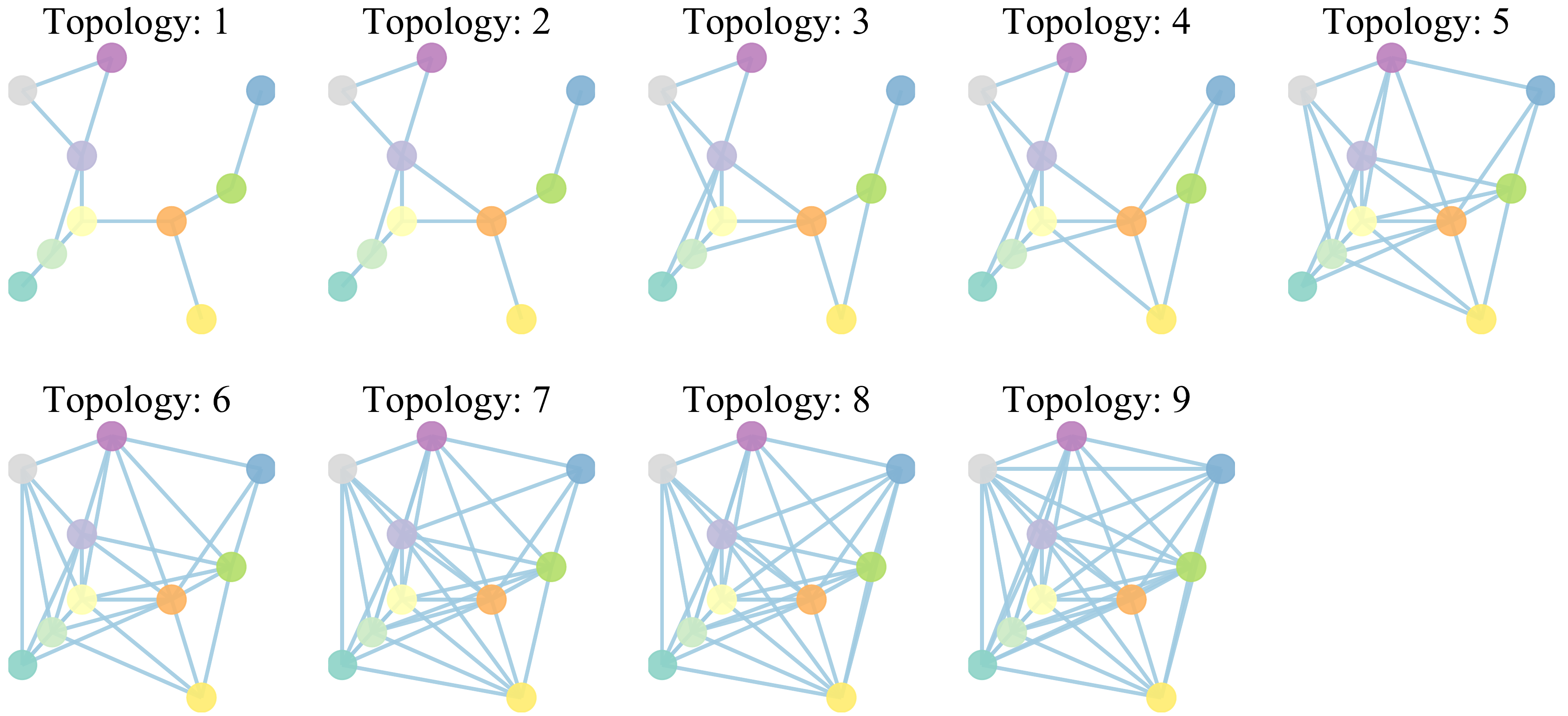}
	\caption{The 9 different topologies used in the sensitivity analysis with respect to topology.}
	\label{case_varytopo}
\end{figure}

Although the computational time of retailers' local calculation is not affected by the change of communication topology, different topologies directly affect the degree of retailers as well as the iteration numbers of the proposed PP-AAC algorithm, resulting in a change in the computational time of the AAC algorithm, which in turn changes the time of the clustering framework. To investigate this trend, we randomly change the communication topology to obtain 9 topologies as shown in Fig. \ref{case_varytopo}.
\vspace{-0.2cm} 

\begin{figure}[h]
	\centering  
	\includegraphics[width=3.5in,center]{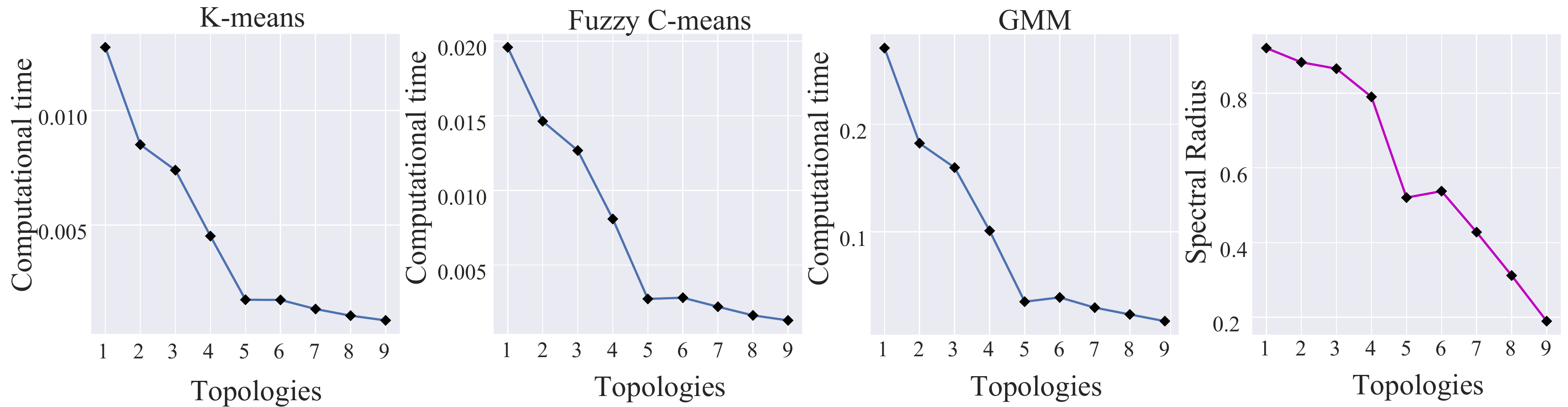}
	\caption{The variation in the average computational times and the spectral radius for different topologies.}
	\label{case_cost_lambda_varytopo}
\end{figure}

Then we measure the total execution time of the AAC algorithm part in the clustering framework for each retailer. Finally, we demonstrate the average time of all retailers when performing the AAC algorithm part in the clustering framework in Fig. \ref{case_cost_lambda_varytopo}. The average $\widetilde{d}_i$ of the retailers and the average iteration numbers of the AAC algorithm under different topologies are also provided in Table \ref{case_degreetimes}. Please note that, the assumption that retailer $i$ cannot receive all the information which its neighbors have received will cause the number of possible communication lines saturate quickly and result in only minor differences between different topologies. Thus we temporarily ignore this assumption to purely demonstrate the variation of cost for different topologies more clearly. 
\vspace{-0.4cm}
\begin{table}[h]
\setlength{\abovecaptionskip}{0pt}
	\renewcommand{\arraystretch}{1.3}
	\caption{The Factors that Affect the Computation Time}
	\label{case_degreetimes}
	\centering
	\footnotesize
	\setlength{\tabcolsep}{1.3mm}{
	\begin{tabular}{l c c c c c c c c c}
	\hline
	\bfseries Topology & 1 & 2 & 3 & 4 & 5 & 6 & 7 & 8 & 9  \\
	\hline
	\bfseries Average $\widetilde{d}_i$ & 1.2 & 1.3 & 1.7 & 1.9 & 2.7 & 2.9 & 3.5 & 3.7 & 4.1 \\
	\bfseries Average iteration number & 121 & 81 & 77 & 25 & 17 & 21 & 14 & 12 & 9  \\
	\hline
	\end{tabular}}
\end{table}
\vspace{-0.5cm}

Theoretically, the average computation overhead of the proposed AAC algorithm part is $\mathcal O(T_aT_c\overline{d}) $, where $\overline{d}$ denotes the average $\widetilde{d}_i$ of all retailers. From Table \ref{case_degreetimes}, we know that although the average $\widetilde{d}_i$ of the retailers increases with the number of the topology map shown in Fig. \ref{case_varytopo}, the increase is much smaller than the decrease of iteration numbers, so the computational time in Fig. \ref{case_cost_lambda_varytopo} is dominated by the iteration numbers. In fact, the worst-case measure of the proposed AAC algorithm's asymptotic convergence rate is proportional to the spectral radius of matrix $(\boldsymbol{W}^\ast-\boldsymbol{J})$, where $\boldsymbol{J}$ is the averaging matrix \cite{4694103}. Since the convergence rate determines the iteration numbers, and the computational time is dominated by the iteration numbers, the trend of the computational time is coincident with the trend of the spectral radius. For verification, we also illustrate the variation in the spectral radius under the different topologies in Fig. \ref{case_cost_lambda_varytopo}. As we can see, the decreasing trends of the computational times and the spectral radius are the same.
\vspace{-0.25cm}
\section{Conclusions}
In this paper, we propose a privacy-preserving distributed clustering framework, which can directly modify the traditional k-means, FCA, and GMM clustering methods and provide privacy-preserving distributed variants. To achieve this, we first performed commonality analysis of the three clustering methods, and pointed out that the key of the clustering framework lies in calculating the summation of the retailers' private information in a fully distributed and privacy-preserving way. Then we developed a PP-AAC algorithm with proven convergence to achieve the summation. Finally, we presented the privacy-preserving distributed clustering framework based on the proposed algorithm with theoretical privacy and complexity analyses. 

The proposed PP-AAC algorithm converges faster than the privacy-preserving AC algorithm and the original AC algorithm. Besides, compared to the original AC algorithm and AAC algorithm, the proposed algorithm is privacy-preserving by introducing the exponentially decaying disturbance with zero-sum property into the shared information. The proposed clustering framework can enable each retailer to obtain the exact residential load pattern identification of all consumers instead of only its own consumers. Thus, this framework can support retailers design better tariff products to attract new users. Meanwhile, the clustering framework not only protects every retailer's privacy, but also greatly reduces the computation overhead of each retailer compared to the centralized method. Moreover, under different communication topologies, the decreasing trends of the PP-AAC part's computational times and the spectral radius are the same.
\vspace{-0.3cm}
\section*{Appendix}
First, we need to prove that $\boldsymbol{W}^\ast$ is doubly stochastic, i.e., that  
\begin{equation}
	\boldsymbol{1}^T\boldsymbol{W}^\ast=\boldsymbol{1}^T, \ \ \boldsymbol{W}^\ast\boldsymbol{1} = \boldsymbol{1} \label{doubly stochastic}
\end{equation}
holds. Define $\boldsymbol{1} \in \Re^{M \times 1} $ as a vector of all ones, then we have:
\begin{align}
	\boldsymbol{1}^T \boldsymbol{W}^\ast  = \boldsymbol{1}^T \boldsymbol{W} + \alpha(\boldsymbol{1}^T\boldsymbol{W} - \boldsymbol{1}^T\boldsymbol{I} ) \notag
\end{align}  
Since $\boldsymbol{W}$ is a doubly stochastic matrix proved in \cite{1440896}, the following holds:
\begin{equation}
	\boldsymbol{1}^T\boldsymbol{W}=\boldsymbol{1}^T, \ \ \boldsymbol{W}\boldsymbol{1} = \boldsymbol{1} \notag
\end{equation}
Substitute $\boldsymbol{1}^T\boldsymbol{W}=\boldsymbol{1}^T$ into $\boldsymbol{1}^T\boldsymbol{W}^\ast $, we obtain
\begin{align}
	\boldsymbol{1}^T \boldsymbol{W}^\ast  = \boldsymbol{1}^T  + \alpha(\boldsymbol{1}^T - \boldsymbol{1}^T ) = \boldsymbol{1}^T  \notag
\end{align}
Similarly, we can obtain $ \boldsymbol{W}^\ast \boldsymbol{1} =\boldsymbol{1}$ with the property that $\boldsymbol{W}\boldsymbol{1} = \boldsymbol{1}$.

Second, define $\boldsymbol{X}^+(t)= [x_1^+(t),...,x_M^+(t)]^T $, $\boldsymbol{X}(t+1)= [x_1(t+1),...,x_M(t+1)]^T $ and $\boldsymbol{\theta}(t) =[\theta_1^+(t),...,\theta_M^+(t)]^T  $. Then we have the matrix form of the proposed PP-AAC algorithm:
\begin{align}
	& \boldsymbol{X}(t+1)  = \boldsymbol{W}^\ast \boldsymbol{X}^+(t) \label{PP AAC algorithm matrix form}\\
	 & \boldsymbol{X}^+(t)= \boldsymbol{X}(t) + \boldsymbol{\theta}(t)  \label{disturbance matrix form} 
\end{align}
In the linear dynamic system in (\ref{PP AAC algorithm matrix form}), as long as $\boldsymbol{W}^\ast$ is doubly stochastic, with the two aforementioned features of $\boldsymbol{\theta}(t)$, the authors in \cite{8685131} proved that (\ref{sum equation}) and (\ref{consensus equation}) hold:
\begin{align}
	& \lim\limits_{t \to \infty }\sum\nolimits_{i=1}^M x_i(t) = \sum\nolimits_{i=1}^M x_i(0) \label{sum equation} \\
	& \lim\limits_{t \to \infty } \mathop{\max}_{\, \,i} \left[x_i(t) \right] - \mathop{\min}_{\, \,i} \left[x_i(t) \right] = 0 \label{consensus equation}
\end{align}
Combining (\ref{sum equation}) and (\ref{consensus equation}) yields (\ref{mean}).         $\quad\quad\quad\quad\quad\quad\qquad \ \blacksquare$

\ifCLASSOPTIONcaptionsoff
  \newind powerage
\fi



\bibliographystyle{IEEEtran}
\bibliography{IEEEabrv,paper}

\begin{thebibliography}{10}
\providecommand{\url}[1]{#1}
\csname url@samestyle\endcsname
\providecommand{\newblock}{\relax}
\providecommand{\bibinfo}[2]{#2}
\providecommand{\BIBentrySTDinterwordspacing}{\spaceskip=0pt\relax}
\providecommand{\BIBentryALTinterwordstretchfactor}{4}
\providecommand{\BIBentryALTinterwordspacing}{\spaceskip=\fontdimen2\font plus
\BIBentryALTinterwordstretchfactor\fontdimen3\font minus
  \fontdimen4\font\relax}
\providecommand{\BIBforeignlanguage}[2]{{%
\expandafter\ifx\csname l@#1\endcsname\relax
\typeout{** WARNING: IEEEtran.bst: No hyphenation pattern has been}%
\typeout{** loaded for the language `#1'. Using the pattern for}%
\typeout{** the default language instead.}%
\else
\language=\csname l@#1\endcsname
\fi
#2}}
\providecommand{\BIBdecl}{\relax}
\BIBdecl

\bibitem{CLEMENTNYNS2011185}
K.~Clement-Nyns, E.~Haesen, and J.~Driesen, ``The impact of vehicle-to-grid on
  the distribution grid,'' \emph{Electr Pow Syst Res.}, vol.~81, no.~1, pp. 185
  -- 192, 2011.

\bibitem{5720519}
H.~{Kanchev}, D.~{Lu}, F.~{Colas}, V.~{Lazarov}, and B.~{Francois}, ``Energy
  management and operational planning of a microgrid with a pv-based active
  generator for smart grid applications,'' \emph{IEEE Trans. Ind. Electron.},
  vol.~58, no.~10, pp. 4583--4592, Oct 2011.

\bibitem{5054916}
P.~{McDaniel} and S.~{McLaughlin}, ``Security and privacy challenges in the
  smart grid,'' \emph{IEEE Secur Priv.}, vol.~7, no.~3, pp. 75--77, May 2009.

\bibitem{4282059}
G.~J. {Tsekouras}, N.~D. {Hatziargyriou}, and E.~N. {Dialynas}, ``Two-stage
  pattern recognition of load curves for classification of electricity
  customers,'' \emph{IEEE Trans. Power Syst.}, vol.~22, no.~3, pp. 1120--1128,
  Aug 2007.

\bibitem{7448460}
Y.~{Wang}, Q.~{Chen}, C.~{Kang}, and Q.~{Xia}, ``Clustering of electricity
  consumption behavior dynamics toward big data applications,'' \emph{IEEE
  Trans. Smart Grid.}, vol.~7, no.~5, pp. 2437--2447, Sep. 2016.

\bibitem{1626400}
G.~{Chicco}, R.~{Napoli}, and F.~{Piglione}, ``Comparisons among clustering
  techniques for electricity customer classification,'' \emph{IEEE Trans. Power
  Syst.}, vol.~21, no.~2, pp. 933--940, May 2006.

\bibitem{1432503}
D.~Z. {Marques}, K.~A. {de Almeida}, A.~M. {de Deus}, A.~R.~G. {da Silva
  Paulo}, and W.~{da Silva Lima}, ``A comparative analysis of neural and fuzzy
  cluster techniques applied to the characterization of electric load in
  substations,'' in \emph{2004 IEEE/PES Transmision and Distribution Conference
  and Exposition}, Nov 2004, pp. 908--913.

\bibitem{6661463}
B.~{Stephen}, A.~J. {Mutanen}, S.~{Galloway}, G.~{Burt}, and
  P.~{Järventausta}, ``Enhanced load profiling for residential network
  customers,'' \emph{IEEE Trans. Power Del.}, vol.~29, no.~1, pp. 88--96, Feb
  2014.

\bibitem{RASANEN20103538}
T.~Räsänen, D.~Voukantsis, H.~Niska, K.~Karatzas, and M.~Kolehmainen,
  ``Data-based method for creating electricity use load profiles using large
  amount of customer-specific hourly measured electricity use data,''
  \emph{Appl. Energy.}, vol.~87, no.~11, pp. 3538 -- 3545, 2010.

\bibitem{LI20161530}
R.~Li, Z.~Wang, C.~Gu, F.~Li, and H.~Wu, ``A novel time-of-use tariff design
  based on gaussian mixture model,'' \emph{Appl. Energy.}, vol. 162, pp. 1530
  -- 1536, 2016.

\bibitem{1056489}
S.~{Lloyd}, ``Least squares quantization in pcm,'' \emph{IEEE Trans. Inf.
  Theory.}, vol.~28, no.~2, pp. 129--137, March 1982.

\bibitem{samet2007privacy}
S.~Samet, A.~Miri, and L.~Orozco-Barbosa, ``Privacy preserving k-means
  clustering in multi-party environment.'' in \emph{SECRYPT}, 2007, pp.
  381--385.

\bibitem{manikandan2018privacy}
V.~Manikandan, V.~Porkodi, A.~S. Mohammed, and M.~Sivaram, ``Privacy preserving
  data mining using threshold based fuzzy cmeans clustering.'' \emph{ICTACT
  Journal on Soft Computing}, vol.~9, no.~1, 2018.

\bibitem{clifton2002tools}
C.~Clifton, M.~Kantarcioglu, J.~Vaidya, X.~Lin, and M.~Y. Zhu, ``Tools for
  privacy preserving distributed data mining,'' \emph{ACM Sigkdd Explorations
  Newsletter}, vol.~4, no.~2, pp. 28--34, 2002.

\bibitem{7903675}
K.~{Xing}, C.~{Hu}, J.~{Yu}, X.~{Cheng}, and F.~{Zhang}, ``Mutual privacy
  preserving $k$ -means clustering in social participatory sensing,''
  \emph{IEEE Trans. Ind. Informat.}, vol.~13, no.~4, pp. 2066--2076, Aug 2017.

\bibitem{leemaqz2017corruption}
K.~L. Leemaqz, S.~X. Lee, and G.~J. McLachlan, ``Corruption-resistant privacy
  preserving distributed em algorithm for model-based clustering,'' in
  \emph{2017 IEEE Trustcom/BigDataSE/ICESS}.\hskip 1em plus 0.5em minus
  0.4em\relax IEEE, 2017, pp. 1082--1089.

\bibitem{4221090}
C.~{Su}, F.~{Bao}, J.~{Zhou}, T.~{Takagi}, and K.~{Sakurai},
  ``Privacy-preserving two-party k-means clustering via secure approximation,''
  in \emph{21st International Conference on Advanced Information Networking and
  Applications Workshops (AINAW'07)}, vol.~1, May 2007, pp. 385--391.

\bibitem{patel2012efficient}
S.~Patel, S.~Garasia, and D.~Jinwala, ``An efficient approach for privacy
  preserving distributed k-means clustering based on shamir’s secret sharing
  scheme,'' in \emph{IFIP International Conference on Trust Management}.\hskip
  1em plus 0.5em minus 0.4em\relax Springer, 2012, pp. 129--141.

\bibitem{7847023}
Z.~{Gheid} and Y.~{Challal}, ``Efficient and privacy-preserving k-means
  clustering for big data mining,'' in \emph{2016 IEEE
  Trustcom/BigDataSE/ISPA}, Aug 2016, pp. 791--798.

\bibitem{meskine2012privacy}
F.~Meskine and S.~N. Bahloul, ``Privacy preserving k-means clustering: a survey
  research.'' \emph{Int. Arab J. Inf. Technol.}, vol.~9, no.~2, pp. 194--200,
  2012.

\bibitem{upmanyu2010efficient}
M.~Upmanyu, A.~M. Namboodiri, K.~Srinathan, and C.~Jawahar, ``Efficient privacy
  preserving k-means clustering,'' in \emph{Pacific-Asia Workshop on
  Intelligence and Security Informatics}.\hskip 1em plus 0.5em minus
  0.4em\relax Springer, 2010, pp. 154--166.

\bibitem{4694103}
T.~C. {Aysal}, B.~N. {Oreshkin}, and M.~J. {Coates}, ``Accelerated distributed
  average consensus via localized node state prediction,'' \emph{IEEE Trans.
  Signal Process.}, vol.~57, no.~4, pp. 1563--1576, April 2009.

\bibitem{8685131}
J.~{He}, L.~{Cai}, P.~{Cheng}, J.~{Pan}, and L.~{Shi}, ``Consensus-based
  data-privacy preserving data aggregation,'' \emph{IEEE Trans. Autom.
  Control.}, pp. 1--1, 2019.

\bibitem{WU2012407}
K.-L. Wu, ``Analysis of parameter selections for fuzzy c-means,'' \emph{Pattern
  Recognit.}, vol.~45, no.~1, pp. 407 -- 415, 2012.

\bibitem{5298967}
R.~{Singh}, B.~C. {Pal}, and R.~A. {Jabr}, ``Statistical representation of
  distribution system loads using gaussian mixture model,'' \emph{IEEE Trans.
  Power Syst.}, vol.~25, no.~1, pp. 29--37, Feb 2010.

\bibitem{1440896}
L.~{Xiao}, S.~{Boyd}, and S.~{Lall}, ``A scheme for robust distributed sensor
  fusion based on average consensus,'' in \emph{IPSN 2005. Fourth International
  Symposium on Information Processing in Sensor Networks, 2005.}, April 2005,
  pp. 63--70.

\bibitem{7465717}
Y.~{Mo} and R.~M. {Murray}, ``Privacy preserving average consensus,''
  \emph{IEEE Trans. Autom. Control.}, vol.~62, no.~2, pp. 753--765, Feb 2017.

\bibitem{MO2010209}
Y.~Mo and B.~Sinopoli, ``Communication complexity and energy efficient
  consensus algorithm,'' \emph{IFAC Proceedings Volumes.}, vol.~43, no.~19, pp.
  209 -- 214, 2010.

\bibitem{Irish}
{Irish Social Science Data Archive}, ``Commission for energy regulation (cer)
  smart metering project.'' \url{http://www.ucd.ie/issda/data/
  commissionforenergyregulationcer/}, 2012.

\bibitem{7579208}
M.~{Sun}, I.~{Konstantelos}, and G.~{Strbac}, ``C-vine copula mixture model for
  clustering of residential electrical load pattern data,'' \emph{IEEE Trans.
  Power Syst.}, vol.~32, no.~3, pp. 2382--2393, May 2017.

\bibitem{LLETI200487}
R.~Lletı́, M.~Ortiz, L.~Sarabia, and M.~Sánchez, ``Selecting variables for
  k-means cluster analysis by using a genetic algorithm that optimises the
  silhouettes,'' \emph{Anal. Chim. Acta.}, vol. 515, no.~1, pp. 87 -- 100,
  2004.

\bibitem{KWEDLO20111613}
``A clustering method combining differential evolution with the k-means
  algorithm,'' \emph{Pattern Recognit. Lett.}, vol.~32, no.~12, pp. 1613 --
  1621, 2011.

\bibitem{internet}
S.~Lai, ``Countries with the fastest internet in the world 2019,'' \emph{ATLAS
  and BOOTS.}, Feb 2019.

\end{thebibliography}
\end{document}